\documentclass[%
 aip,
 amsmath,amssymb,
 reprint,%
]{revtex4-1}

\usepackage{graphicx}
\usepackage{dcolumn}
\usepackage{bm}

\usepackage[utf8]{inputenc}
\usepackage[T1]{fontenc}
\usepackage{mathptmx}

\begin{document}

\preprint{AIP/123-QED}

\title{Local characterization of a heavy-fermion superconductor via sub-Kelvin magnetic force microscopy}

\author{Dirk Wulferding}
\altaffiliation{Contributed equally to this work.}
\affiliation{Center for Correlated Electron Systems, Institute for Basic Science (IBS), Seoul 08826, Korea}
\affiliation{Department of Physics and Astronomy, Seoul National University (SNU), Seoul 08826, Korea}

\author{Geunyong Kim}
\altaffiliation{Contributed equally to this work.}
\affiliation{Max Planck POSTECH Center for Complex Phase Materials, Pohang University of Science and Technology, Pohang 37673, Korea}
\affiliation{Department of Physics, Pohang University of Science and Technology, Pohang 37673, Korea}

\author{Hoon Kim}
\affiliation{Department of Physics, Pohang University of Science and Technology, Pohang 37673, Korea}
\affiliation{Center for Artificial Low Dimensional Electronic Systems, Institute for Basic Science, Pohang 37673, Korea}

\author{Ilkyu Yang}
\affiliation{Department of Physics, Pohang University of Science and Technology, Pohang 37673, Korea}
\affiliation{Center for Artificial Low Dimensional Electronic Systems, Institute for Basic Science, Pohang 37673, Korea}

\author{E. D. Bauer}
\affiliation{Los Alamos National Laboratory, MPA-CMMS, Los Alamos, NM 87545, USA}

\author{F. Ronning}
\affiliation{Los Alamos National Laboratory, MPA-CMMS, Los Alamos, NM 87545, USA}

\author{R. Movshovich}
\affiliation{Los Alamos National Laboratory, MPA-CMMS, Los Alamos, NM 87545, USA}

\author{Jeehoon Kim}
\email[]{Corresponding author: jeehoon@postech.ac.kr}
\affiliation{Max Planck POSTECH Center for Complex Phase Materials, Pohang University of Science and Technology, Pohang 37673, Korea}
\affiliation{Department of Physics, Pohang University of Science and Technology, Pohang 37673, Korea}

\date{\today}

\begin{abstract}
Using magnetic force microscopy operating at sub-Kelvin temperatures we characterize the heavy-fermion superconductor CeCoIn$_5$. We pinpoint the absolute London penetration depth $\lambda(0) = 435 \pm 20$ nm and report its temperature dependence, which is closely linked to the symmetry of the superconducting gap. In addition, we directly measure the pinning force of individual Abrikosov vortices and estimate the critical current density $j_c = 9 \times 10^4$ A/cm$^2$. In contrast to the related, well-established tunnel diode oscillator technique, our method is capable of resolving inhomogeneities $locally$ on the micrometer-scale at ultra-low temperature.
\end{abstract}

\maketitle

To date, there is a continuing dispute over the mechanisms that lead to superconductivity in various unconventional and exotic materials, ranging from cuprates~\cite{keimer-15} to Fe-based compounds~\cite{hosono-15} to heavy-fermion systems.~\cite{thalmeier-05} Hence, a goal is to unambiguously understand the nature of superconductivity in these fundamentally different classes of materials. A key ingredient towards reaching this goal is the symmetry of the superconducting gap, which can help to reveal the pairing mechanism.~\cite{hirschfeld-16} So far, various approaches have been employed to gain insight into the gap symmetry,~\cite{tsuei-00} from fully bulk methods -- specific heat,~\cite{yonezawa-17} nuclear magnetic resonance,~\cite{nakai-10} tunnel diode oscillator technique~\cite{shang-20} -- to local and/or surface sensitive probes -- scanning tunneling spectroscopy,~\cite{hoffman-11} angle-resolved photoemission spectroscopy,~\cite{huang-12} Raman spectroscopy.~\cite{gallais-16} Nevertheless, results obtained from different methods can lead to conflicting conclusions, as each method is plagued by its own drawbacks that might obscure the true gap symmetry in different ways. Consequently, it is of importance to have a variety of complementary techniques at hand.

Magnetic force microscopy (MFM) has been shown to be an apt method to measure the superfluid density locally and as a function of temperature in the Fe-pnictide Ba(Fe,Co)$_2$As$_2$.~\cite{luan-10} Here we employ our home-built MFM operating at a base temperature of 500 mK for a local characterization of CeCoIn$_5$~\cite{petrovic-01} in the sub-Kelvin regime. Through a comparative approach we determine the London penetration depth $\lambda$ with high accuracy, which in previous reports varied by a factor of 2. The highly-resolved $\lambda(T)$ evidences $d$-wave behavior in accordance with previous reports, proving the validity of our method. In addition, we extract the absolute value of the Abrikosov vortex pinning force in this heavy-fermion superconductor, and discuss the robustness of its superconducting phase against thermal fluctuations. MFM thus offers a fast, simple, and straightforward technique for the local characterization of exotic, low-$T_c$ superconductors, that can be realized in an inexpensive table-top setup.

A single crystal of CeCoIn$_5$ ($T_c=2.25$ K) was grown via the self-flux method.~\cite{petrovic-01} Prior to MFM measurements, the sample was oriented with respect to its $c$ axis via X-ray diffraction and subsequently mechanically polished to obtain a fresh surface within the $ab$ plane with a roughness as small as 10 nm. The resulting dimensions of the measured specimen were 1.5 mm $\times$ 1.5 mm $\times$ 0.5 mm. MFM measurements were performed using a home-built $^3$He MFM probe with a temperature range of 500 mK -- 300 K.~\cite{yang-16} All experiments were carried out using the same commercially available cobalt-alloy coated pyramidal silicon tip with a tip radius of curvature < 50 nm (PPP-MFMR from NANOSENSORS$\texttrademark$). The force gradient $\partial F / \partial z$ is obtained from the measured frequency shift $\Delta f$ via $\frac{\partial F}{\partial z} = -2k \frac{\Delta f}{f_0}$, where $k = 2.8$ N and $f_0 = 78912.3$ Hz are the force constant and the bare resonance frequency of the MFM cantilever, respectively.

\begin{figure}
\label{figure1}
\centering
\includegraphics[width=8cm]{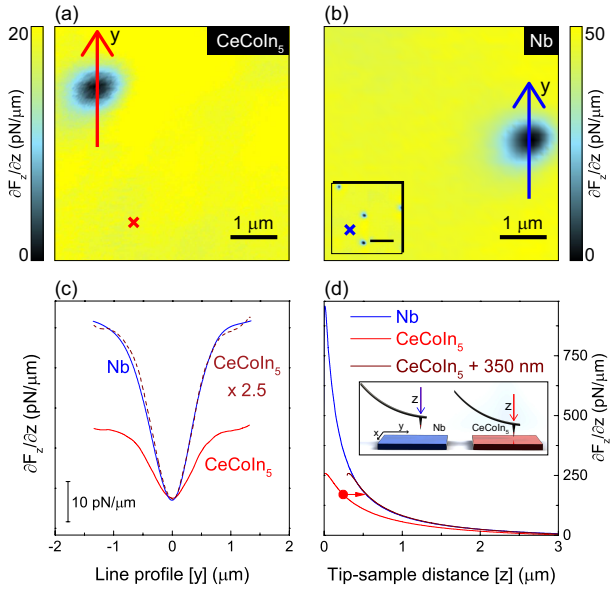}
\caption{(a) 5 $\mu$m $\times$ 5 $\mu$m MFM image of CeCoIn$_5$ obtained in the superconducting state at $T = 0.5$ K with a tip-sample lift height of 300 nm. (b) 5 $\mu$m $\times$ 5 $\mu$m MFM image of the Nb film obtained in the superconducting state at $T = 4.2$ K with a tip-sample lift height of 200 nm. Inset: 15 $\mu$m $\times$ 15 $\mu$m scan frame, the scale bar corresponds to 5 $\mu$m. (c) Line profiles through the vortices as indicated by the arrows in (a) and (b). (d) As-measured Meissner force curves on CeCoIn$_5$ (red) and the Nb film (blue), as well as the shifted Meissner force curve for CeCoIn$_5$ (dark red). The measurement positions are marked by crosses in (a) and (b).}
\end{figure}

We measure the London penetration depth $\lambda$ through a comparative method,~\cite{luan-10, kim-12} which requires the simultaneous measurement of a well-characterized standard sample, i.e., a 300 nm thin film of Nb (see Ref.~\cite{nazaretski-09} for a detailed characterization of the standard sample used in our study). The penetration depth of this thin film was previously determined as $\lambda_{\mathrm{Nb}} = (105 \pm 5)$ nm at 4.3 K.~\cite{nazaretski-09} This value is larger than that of bulk, clean Nb ($\lambda \approx 39$ nm), which is owed to a certain amount of disorder and defects in the thin film. In Figs. 1(a) and 1(b) we show MFM images of CeCoIn$_5$ and the Nb film in their superconducting state (at 0.5 K and 4.2 K, respectively). We observe one dark vortex in each image on an otherwise homogeneous background, indicating a homogenous superfluid density $\rho$. These vortices originate from the earth's magnetic stray field as well as from stray fields of the instrument and the laboratory environment. The samples trap these stray fields in the form of magnetic flux quanta once they enter their superconducting state. In order to minimize the influence from vortices on our local probe experiments, we apply a magnetic field which mostly compensates the stray fields.~\cite{yang-16} We can now estimate the strength of the remaining uncompensated stray field $H_{\mathrm{rem}}$ by considering the number of vortices $N$ per scanned area $A$, i.e., $H_{\mathrm{rem}} = \frac{N \Phi_0}{A}$. Here, $\Phi_0$ is the magnetic flux quantum. The inset in Fig. 1(b) shows the appearance of 4 vortices on a large scan area of 15 $\mu$m $\times$ 15 $\mu$m. This corresponds to $H_{\mathrm{rem}} = 0.4$ Oe, which is smaller than the accuracy of our magnet with a resolution of 1 Oe.

A comparison between line profiles through the vortices, indicated by the arrows in Figs. 1(a) and 1(b), is shown in Fig. 1(c). The near-perfect coincidence of the scaled CeCoIn$_5$ cut (dashed dark-red curve) with the Nb film cut (blue curve) highlights vortices as the common origin for the circular shaped features in the MFM images of CeCoIn$_5$ and the Nb film. The smaller absolute magnitude of the vortex line profile for CeCoIn$_5$ compared to that of the Nb film indicates that CeCoIn$_5$ has a larger penetration depth $\lambda$ than the Nb film, as the repulsive Meissner force on the tip at similar lift heights is weaker. Having access to the absolute value of $\lambda$ is of major importance in understanding superconductivity. This value is directly related to the density of Cooper pairs, and through the Uemura relation,~\cite{uemura-88} $\lambda^{-2}(0) \sim T_c$ can be found in various classes of superconductors. In combination with the coherence length $\xi$, $\lambda$ also yields the Ginzburg-Landau parameter $\kappa$ for a classification into type-I or type-II superconductors, as well as the Ginzburg number $Gi$, which is essential for quantifying thermal fluctuations within the superconducting state.~\cite{tinkham, blatter-94} In order to extract the absolute value of $\lambda$, we slowly lower the magnetic tip in $z$ direction towards the surface of superconducting CeCoIn$_5$ at a fixed lateral position several $\mu$m away from Abrikosov vortices [marked by '$\times$' in Figs. 1(a) and (b)]. This results in the red Meissner force curve shown in Fig. 1(d). The strength of the Meissner force depends on the magnetic penetration depth, as well as on experimental details such as the tip magnetization and the tip geometry. We can account for these latter effects by directly comparing the Meissner force curve to that of our Nb film reference sample. Recall that we use the same experimental conditions via our multi-sample stage.~\cite{yang-16} In this case, any difference between the Meissner force curves is solely due to a difference in $\lambda$. Therefore, the penetration depth of CeCoIn$_5$ can be obtained simply by vertically shifting its Meissner force curve to coincide with that of the Nb film [here, a shift by 350 nm along the $z$ direction leads to the collapse of both curves; see red arrow in Fig. 1(d)]. Our method yields a penetration depth at $T = 500$ mK of $\lambda = \lambda_{\mathrm{Nb}}$ + 350 nm = 455 nm with an accuracy of $\Delta \lambda = 20$ nm. Considering the temperature dependence of $\lambda$ (as discussed below), the absolute penetration depth at $T = 0$ K amounts to $\lambda(0) = 435 \pm 20$ nm. This value is in-between previous indirect measurements, which found the zero-temperature penetration depth ranging from $\lambda_{\mathrm{impedance}} = 281$ nm [Ref. \cite{ozcan-03}] to $\lambda_{\mu \mathrm{SR}} = 550$ nm.~\cite{higemoto-02} We emphasize that the absolute magnetic penetration depth of CeCoIn$_5$ is pinned down with our method at $T$ = 0 with high accuracy of about 5\%.
Using the reported values for the coherence length \mbox{$\xi = 5.0$ nm} [Ref. \cite{debeer-06, zhou-13}] and the anisotropy parameter $\gamma = \lambda_{c} / \lambda_{ab} = 1.25$,~\cite{howald-13} we calculate $Gi = \gamma^2/2 \cdot \left[ \frac{\mu_0 k_B T_c}{4 \pi B_c^2(0) \xi^3} \right]^2 = 4 \cdot 10^{-6}$. Here, \mbox{$B_c(0) = \phi_0 / [2 \sqrt{2} \pi \lambda(0) \xi(0)] = 1070$ Oe}. The obtained Ginzburg number is slightly larger than typical values of conventional superconductors ($Gi \sim 10^{-7}$) [Ref. \cite{blatter-94}] and much smaller than that of cuprates or pnictides ($Gi \sim 10^{-1} - 10^{-2}$),~\cite{blatter-94, kim-pnictide} highlighting that superconductivity in CeCoIn$_5$ is very robust against thermal fluctuations. It narrows the theoretical width of the superconducting transition down to $\Delta T_c = Gi \cdot T_c = 25$ $\mu$K.

Through our measurement of $\lambda(0)$ we can also deduce the absolute superfluid density $\rho_0$ at $T = 0$ K in a heavy fermion superconductor via $\lambda(0)^2 = \frac{m^*}{\mu_0 \rho_0 e^2}$. With $m^* \approx 49 m_e$ [Ref. \cite{settai-01, mccollam-05}] being the effective electron mass in CeCoIn$_5$, $\rho_0 = 7.3 \cdot 10^{27}$ m$^{-3}$. Please note that this value is a rough estimate based on a simple one-band picture, where the actual anisotropy of $m^*$ was not taken into account. Nevertheless, it yields the correct order of magnitude for the superfluid density in CeCoIn$_5$ which is comparable to densities reported in cuprates~\cite{xi-92} and iron-based superconductors~\cite{sun-17} (ranging from $2 - 5 \times 10^{27}$ m$^{-3}$).

\begin{figure}
\label{figure2}
\centering
\includegraphics[width=8cm]{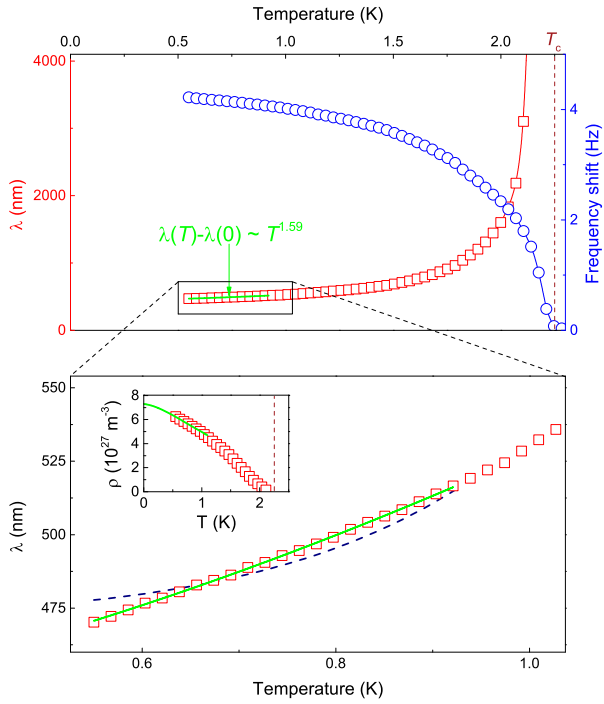}
\caption{Upper panel: Temperature-dependent frequency shift of the oscillating MFM tip at a fixed, finite tip-sample distance (blue circles) together with the converted temperature-dependent absolute value of the penetration depth $\lambda$ (red squares). Lower panel: zoom-in of the low temperature range together with a power-law fit (green solid line) and a Gorter-Casimir fit (dark blue dashed curve) up to $1/3 T_c$. The inset plots the absolute superfluid density $\rho$ as a function of temperature (red squares) together with its fit (green solid line).}
\end{figure}

We now turn to the gap symmetry measured locally via MFM. As was demonstrated in Fig. 1(d), the shape of the Meissner force curve does not reveal any information about the gap symmetry, as both $s$-wave and $d$-wave superconductors result in the same $z$-dependence of the Meissner force. Instead, we investigate the temperature evolution of $\lambda$. The magnetic tip is placed at a finite, fixed distance of 500 nm above the surface of CeCoIn$_5$ at $T = 500$ mK. This yields a constant Meissner force, detected as a constant, positive shift in the tip's resonance frequency. We steadily heat up the sample to 2.5 K. As the superconductor approaches $T_c$ the superfluid density continuously decreases, which leads to a reduction in the Meissner force. Hence the tip's frequency shift relaxes back to 0, i.e., the characteristic, unperturbed resonance frequency $f_0$ is recovered for $T > T_c$, as seen in Fig. 2, empty blue circles. Using the comparative method discussed in Fig. 1, we convert the frequency shift into absolute values of the penetration depth, which allows us to plot $\lambda$ as a function of temperature (Fig. 2, empty red squares). It should be noted that the eigenfrequency of a cantilever shows a temperature dependence, mainly caused by a thermally-induced change of exchange gas pressure inside the MFM probe; this unwanted frequency shift is especially relevant when covering wide temperature ranges, and can be compensated by monitoring $\Delta f(T)$ far away from the superconducting surface. However, on a scale of a few K this effect is minute, and especially negligible at sub-Kelvin temperatures, allowing a precise measurement of $\lambda(T)$ in heavy fermion superconductors. Additionally, with increasing temperature the superconducting gap $\Delta$ starts to diminish, while at low temperatures ($T<1/3T_c$) it can be considered nearly temperature-independent. For these reasons, generally $\lambda(T)$ is only studied in the low-temperature regime $T<1/3T_c$.~\cite{prozorov-06} In a $d$-wave superconductor such as CeCoIn$_5$,~\cite{izawa-01} the penetration depth in the low-temperature limit follows a power-law behavior as a function of the temperature.~\cite{prozorov-06} We apply a fit according to $\lambda(T) - \lambda(0) \sim T^n$ to the data up to $T \approx 0.9$ K, as indicated by the solid green line in the upper panel. The fit yields a zero-temperature penetration depth of $\lambda(0) = (435 \pm 20)$ nm and an exponent of $n=1.59 \pm 0.07$. This exponent is well-consistent with the scenario of nodal $d$-wave superconductivity, and comparable to that of cuprates with $d$-wave pairing symmetry, where the exponent ranges between 1 and 2.~\cite{prozorov-06} Previous studies performed with a tunnel diode oscillator technique, i.e., a bulk method, found evidence for a $d$-wave superconducting gap with exponents between $n = 1.43$ and $n = 1.57$.~\cite{chia-03, ozcan-03} The lower panel zooms in onto the low temperature regime, highlighting an excellent match between data and fit (the coefficient of determination for the fit is $R^2 = 0.9995$). As a comparison, we also applied a BCS-type fit in the form of a low-temperature Gorter-Casimir approximation,~\cite{tinkham} $\lambda(T) = \frac{\lambda(0)}{\sqrt{1-(T/T_c)^4}}$, which would be in accordance with a conventional $s$-wave behavior. For our fit, both $\lambda(0)$ and $T_c$ are free parameters. As can be seen from the dashed dark blue curve, the fit clearly deviates from the experimental data ($R^2 = 0.9618$). In the inset we plot the superfluid density $\rho(T) \sim \lambda(T)^{-2}$ to show its demise with increasing temperature.

\begin{figure}
\label{figure3}
\centering
\includegraphics[width=8cm]{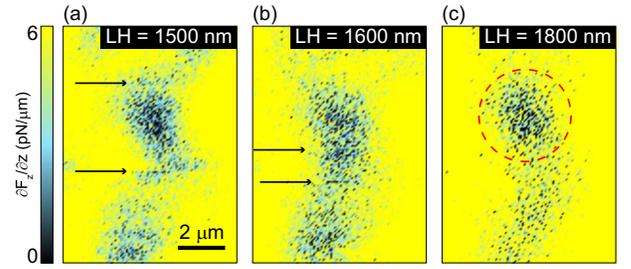}
\caption{MFM images obtained at $T=0.5$ K at various tip lift heights (LH): (a) 1500 nm, (b) 1600 nm, and (c) 1800 nm. Arrows denote artifacts due to tip manipulation. The dashed circle in (c) marks the unperturbed vortex.}
\end{figure}

In order to extract the pinning force of individual vortices we successively image the sample within the superconducting state at various tip-sample distances.~\cite{straver-08} Fig. 3 presents MFM images taken at $T = 500$ mK with increasing lift heights. Fig. 3(a) with a lift height of 1500 nm clearly shows several artifacts (marked by arrows). These are evidence for vortex manipulation -- and thus, unpinning -- by the magnetic moment of the tip. At a distance of 1600 nm [Fig. 3(b)] these artifacts are strongly reduced, and they completely vanish for lift heights of 1800 nm [Fig. 3(c)] and larger. Artifact-free vortex images can therefore only be obtained at very large tip-sample distances. Hence, the detected MFM signal is very weak, resulting in noisy MFM images as seen in Fig. 3. We estimate the magnetic moment per unit length of the tip $m_{\mathrm{tip}} = (8.5 \pm 0.3)$ nA$\cdot$m through a simple monopole-monopole approximation,~\cite{auslander-09} and obtain the force between tip and vortex $F_{\mathrm{tv}} = \frac{m_{\mathrm{tip}}\Phi_0}{2 \pi} \times \frac{1}{[z+\lambda(T)]^2}$, which would be compensated by the intrinsic pinning force of individual vortices in CeCoIn$_5$ in the case of static, artifact-free images. Here, $z$ corresponds to the tip-sample distance at which vortices start to unpin and artifacts appear. Considering a distance between $z=1600$ nm and 1800 nm, we find a pinning force of $F_{\mathrm{p}} = (0.61 \pm 0.08)$ pN at 500 mK. In contrast, the vortex in Fig. 1(a), imaged at a lift height of 300 nm, evidences no artifacts due to tip-vortex interaction, suggesting a pinning force stronger than 5 pN. Here we mention that in our study only 4 out of totally 67 MFM images of superconducting CeCoIn$_5$ featured clear vortices at small tip-sample distances. This observation suggests that the vortex imaged in Fig. 1(a) is strongly pinned by a random, local defect, and that 0.61 pN is the intrinsic pinning force of CeCoIn$_5$ in the absence of local defects. The pinning force in cuprate and Fe-based superconductors is typically 1-2 orders of magnitude larger.~\cite{auslander-09, zhang-15, kim-unpub} The remarkably small intrinsic pinning force in CeCoIn$_5$ reflects that it is within the superclean regime, i.e., $\epsilon_F / \Delta \ll \ell / \xi$ ($\epsilon_F$ = Fermi energy, $\ell$ = quasiparticle mean free path).~\cite{kasahara-05} From an applied point of view, however, small pinning forces are rather detrimental as strong vortex pinning can increase critical current densities considerably. We can roughly estimate the critical current density $j_c$(0.5K) $\approx 9 \cdot 10^4$ A/cm$^2$ via the relation~\cite{blatter-94} $\frac{\Phi_0}{c} \vec{j_c} \times \vec{n} = \vec{F_{\mathrm{p}}}$, with $F_{\mathrm{p}} = 0.61$ pN. $\vec{n}$ denotes the unit vector along the vortex, and $c$ is the speed of light. Our simple approximation agrees well with $j_c = 7 \cdot 10^4$ A/cm$^2$ extracted from earlier ultrasound velocity measurements.~\cite{watanabe-04} The good agreement between global and local measurements suggests that CeCoIn$_5$ is a clean and homogeneous heavy fermion superconductor.

Finally, we mention two additional MFM-related experimental fingerprints to directly image the gap anisotropy, albeit with limited relevance to heavy-fermion superconductors: firstly, using a weakly magnetized tip, a strongly pinned vortex could be imaged at small tip-sample distances (but large enough to avoid any vortex manipulation) with a sufficiently high lateral resolution, that might reveal vortex shape anisotropy.~\cite{zhou-13} Secondly, the tip-induced vortex manipulation can reveal the existence of line nodes through an anisotropic wiggling motion, as previously shown in cuprates.~\cite{auslander-09} However, whether these methods are applicable in the superclean limit of CeCoIn$_5$ with exceptionally weak pinning remains an open issue to be addressed in future studies.

We presented a comprehensive local characterization of the superconducting properties for the exotic, heavy-fermion system CeCoIn$_5$ using magnetic force microscopy at ultra-low temperatures. We extracted the absolute value for the penetration depth $\lambda(0)=435 \pm 20$ nm together with the Ginzburg number $Gi$=4$\times$10$^{-6}$, and reported the pinning force $F_{\mathrm{p}}=0.61$ pN, as well as the critical current density $j_c=9 \times 10^4$ A/cm$^2$ at 500 mK. We also demonstrated the possibility of locally probing the superconducting gap symmetry at sub-Kelvin ($^3$He) temperatures. The $d$-wave character of CeCoIn$_5$, with a power-law behavior of $\lambda(T)$ and an exponent of $n=1.59$ is in good agreement with previous reports. Our instrument opens the door to investigate not only the nature of superconductivity in exotic compounds with low transition temperatures, but particularly in samples with intrinsic anisotropies or competing orders~\cite{wulferding-15, kamlapure-17} that might lead to a local variation of the gap symmetry. It furthermore allows us to directly study the effect of impurities and defects on superconductivity on a micrometer scale.

\begin{acknowledgments}
This work was supported by the Institute for Basic Science (IBS) (Grant No. IBS-R014-D1 and Grant No. IBS-R009-Y3). G.K. and J.K. were supported by the Ministry of Education, Science, and Technology (No. NRF-2016K1A4A01922028, NRF-2018R1A5A6075964, and NRF-2019R1A2C2090356). Work at LANL was supported by the U.S. Department of Energy, Basic Energy Sciences, Division of Materials Sciences and Engineering.
\end{acknowledgments}

\section{DATA AVAILABILITY}
The data that support the findings of this study are available from the corresponding author upon reasonable request.

\end{document}